\documentclass{aa}  

\usepackage{graphicx}
\usepackage{txfonts}
\usepackage[
  breaklinks = true,
   colorlinks = true,
  urlcolor   = blue,
  citecolor  = blue,
  linkcolor  = red, ]{hyperref}
\graphicspath{ {./figures/} }

\begin{document} 

   \title{Generic low-atmosphere signatures of swirled-anemone jets}
 \titlerunning{Low-atmospheric signatures of swirled-anemone jets}
   \author{Reetika Joshi
          \inst{1,2}
          \and
      Guillaume Aulanier \inst{3,1}
     \and 
         Alice Radcliffe \inst{3}
          \and
          Luc Rouppe van der Voort \inst{1,2}
          \and
          Etienne Pariat \inst{3}
          \and
          Daniel Nóbrega-Siverio \inst{4,5,1,2}
          \and
          Brigitte Schmieder \inst{6,7,8}
          }

   \institute{Rosseland Centre for Solar Physics, University of Oslo, P.O. Box 1029 Blindern, N-0315 Oslo, Norway\\
   \email{reetika.joshi@astro.uio.no}
    \and
    Institute of Theoretical Astrophysics, University of Oslo, P.O. Box 1029 Blindern, N-0315 Oslo, Norway  
    \and
    Sorbonne Universit\'e, École Polytechnique, Institut Polytechnique de Paris, Observatoire de Paris - PSL, CNRS, Laboratoire de physique des plasmas (LPP), 4 place Jussieu, F-75005 Paris, France
         \and
    Instituto de Astrof\'{\i}sica de Canarias, E-38200 La Laguna, Tenerife, Spain 
    \and
    Departamento de Astrof\'{\i}sica, Universidad de La Laguna, E-38206 La Laguna, Tenerife, Spain 
         \and
    LESIA, Observatoire de Paris, Universit\'e PSL, CNRS, Sorbonne
    Universit\'e, Universit\'e de  Paris,
    5 place Jules
    Janssen, F-92195 Meudon, France
    \and
    Centre for mathematical Plasma Astrophysics, Dept. of Mathematics, KU Leuven, 3001 Leuven, Belgium
     \and
    School of Physics and Astronomy, University of Glasgow, Glasgow G12 8QQ, UK
             }


 
  \abstract
   {Solar jets are collimated plasma flows moving along magnetic field lines 
   and accelerated at low altitude following magnetic reconnection. Several of them originate from anemone-shaped low--lying arcades and the most impulsive ones tend to be relatively wider and display untwisting motions.} 
   {We aim to establish typical behaviours and observational signatures in the low atmosphere that can occur in response to the coronal development of such impulsive jets.}
   {We analysed an observed solar jet associated with a circular flare ribbon, using high-resolution observations from SST coordinated with IRIS and SDO. 
   We related specifically--identified features with those developing in a generic 3D line-tied numerical simulation of reconnection--driven jets, performed with the ARMS code.}
  {We identified three features in the SST observations: 
   the formation of a hook along the circular ribbon, the gradual widening of the jet through the apparent displacement of its kinked edge towards --and not away-- from the presumed reconnection site, and the falling back of some of the jet plasma towards a footpoint offset from that of the jet itself. 
  The 3D numerical simulation naturally accounts for these features which were not imposed a priori. Our analyses allow to interpret them in the context of the 3D geometry of the asymmetric swirled anemone loops and their sequences of reconnection with ambient coronal loops.}
   {Given the relatively-simple conditions in which the observed jet occurred, together with the generic nature of the simulation that comprised minimum assumptions, we predict that the specific features that we identified and interpreted are probably typical of every impulsive jet.} 


   \maketitle
%

\section{Introduction}

Solar jets are collimated plasma ejections observed throughout the solar atmosphere.
They  have been extensively studied in terms of their morphology, dynamic characteristics, and driving mechanisms since their first detection in the X-ray emission of coronal jets by the Soft X-ray Telescope aboard the Yohkoh satellite in the early 1990s \citep{Shibata1992,Schmieder1995,Shibata1996}. Jet footpoints are observed  along magnetic field inversion lines in case of flux cancellation process \citep{Chifor2008,Sterling2018} or in regions of high current density layers with a high squashing factor (Q) called Quasi-Separatrix layers  (QSL) \citep{Demoulin1996, Joshi2017b}. Presently, several reviews discuss the state of art of jets using several space satellites and ground-based observations \citep{Raouafi2016,Shen2021,Schmieder2022}.
 
 Two morphological different types of solar jets, ``straight" and ``blow-out" were introduced by \citet{Moore2010}, using the Hinode/XRT observations. In these observations, it has been found that in some X-ray jets the jet spire remained narrow and the base appeared dim and inert during the jet eruption process. In contrary to these narrow spire jets, there were other X-ray jets which evolve with wider spire and during the eruption, the spire width becomes comparable to the bright and active jet base.
 It is believed that the narrow-spire straight jets are generated following the standard jet model proposed by \citet{Heyvaerts1977} and \citet{Shibata1992}, hence called ``standard jets”. However, in the case of broad-spire jets, the eruption blows out the involved emerging bipole magnetic field and 
 with it, cool chromospheric material along with the hot jet ejects, 
 hence named as ``blowout" jets \citep[for more details:][]{Raouafi2016}. 
\citet{Pariat2015} showed that the formation of either standard jet or blowout jet (which can also be called helical jets) could be governed by differences in the reconnection mode and rate and by the driving process of the magnetic system.
 
Jet-like events have been detected in almost all wavelengths available to observers. They are called surges when they are observed in absorption in chromospheric spectral lines (H$\alpha$)
and are frequently associated with hot jets \citep[see, e.g.,][]{Shibata1995,Mandrini2002,Uddin2012,Joshi2020,Joshi2021Balmer}. Several theoretical models and numerical experiments explain the ejection of cool plasma during the jet eruption phase  by the rise of chromospheric plasma around the jet in the emerging magnetic flux model \citep{Moreno2008, Moreno2013, Nobrega2016, Nobrega2017}.


There is a general consensus on magnetic reconnection being a necessary process for the generation of coronal jets. Magnetic reconnection between an emerging photospheric magnetic field and a pre-existing magnetic field can occur at different spatial scales. Reconnection happening in the solar corona produces high velocity hot jets (10$^6$ K) observable in X-ray or EUV \citep{Shimojo2000}, whereas cooler H$\alpha$ surges can be produced by reconnection occurring at the chromospheric layer \citep{Shibata2007} or by slingshot effect next to the hot jets \citep{Yokoyama1996}.  When reconnection takes place between a small emerging bipole with an ambient magnetic field of opposite polarity, it gives birth to an anemone-shaped jet, where an opposite polarity connects to the ambient magnetic field in a way forming a fan like shape similar to a sea anemone \citep{Shibata2007,Nishizuka2011, Singh2012}.

Magnetic reconnection as a generator for the solar jets has been also explored in several numerical simulations, such as: \citet{Yokoyama1995}, \citet{Moreno2008}, and \citet{Torok2009} 
performed numerical simulations based on
the magnetic reconnection model to reproduce coronal X-ray
jets, which successfully demonstrate the connection between
jets and magnetic reconnection. 
At the reconnection site the plasma may  be accelerated by the propagation of magnetic tension through Alfvén waves. In the absence of strong shear or twist in the system (typical of 2D simulations), this wave is the direct consequence of the reconnection-jet often referred as the slingshot effect \citep{Nobrega2017}. When the initially closed system comprises
  twisted fields, a  torsional Alfvén wave is induced in the newly reconnected field line due to the force imbalance  between the twisted and the untwisted parts \citep{Pariat2009, Torok2009, Wyper2016}.

In addition to these, the jet plasma can be accelerated by ``siphon flow'' due to the pressure difference between the top of the newly open field lines and its closed bottom part \citep{Scott2022} and by "evaporation" upflow  due to the additional energy deposition induced by reconnection near the jet footpoint. This increases the internal energy at the bottom of the newly reconnected jet loops forming a difference in pressure gradient leads inducing the upflows \citep{Shimojo2001}.

The main aim for our present study is to probe the signatures of reconnection in a wide and impulsive jet. Those wide and impulsive jets often show twisting features while going up to evacuate. We address a few questions, such as: can we see the observational signatures of these jets in the lower corona or what is the response in the low corona to these kind of jets, which are very common jets. 
We present a jet event with coordinated  observations, which reveal that the jet  was originated from a swirled anemone, which is the perfect signature of large-scale non potential field as a jet progenitor. We choose this simple case event with only one extended magnetic polarity and no mini-filament signatures at the jet base, so the shear is distributed all over the polarity inversion line (PIL). 
We compare our observational signatures with a theoretical model to address a few questions, for instance: how a jet develops? 
What are the responses to the low solar atmosphere to that jet eruption? Among these responses, are some of the puzzling observed features actually generic and not atypical? We aim for a simple model, where the whole parasitic polarity is rotating and this leads us to the model given by \citet{Pariat2009}. Our observational event is not fully axis-symmetric like in this model, but it appears to be slightly asymmetric and inclined. So we keep to the same idea from \citet{Pariat2009}, but use an inclined version of the model well explained in  \citet{Pariat2015}. 

The paper is organised as: an overview of used instruments and observational features of the jet using multi-instruments are presented in Sect. \ref{observations}, an introduction to the used model and comparison with observed features are explained in Sect. \ref{model}. The results are discussed in Sect. \ref{results}.

   \begin{figure*}[ht!]
   \centering
   \includegraphics[width=\textwidth]{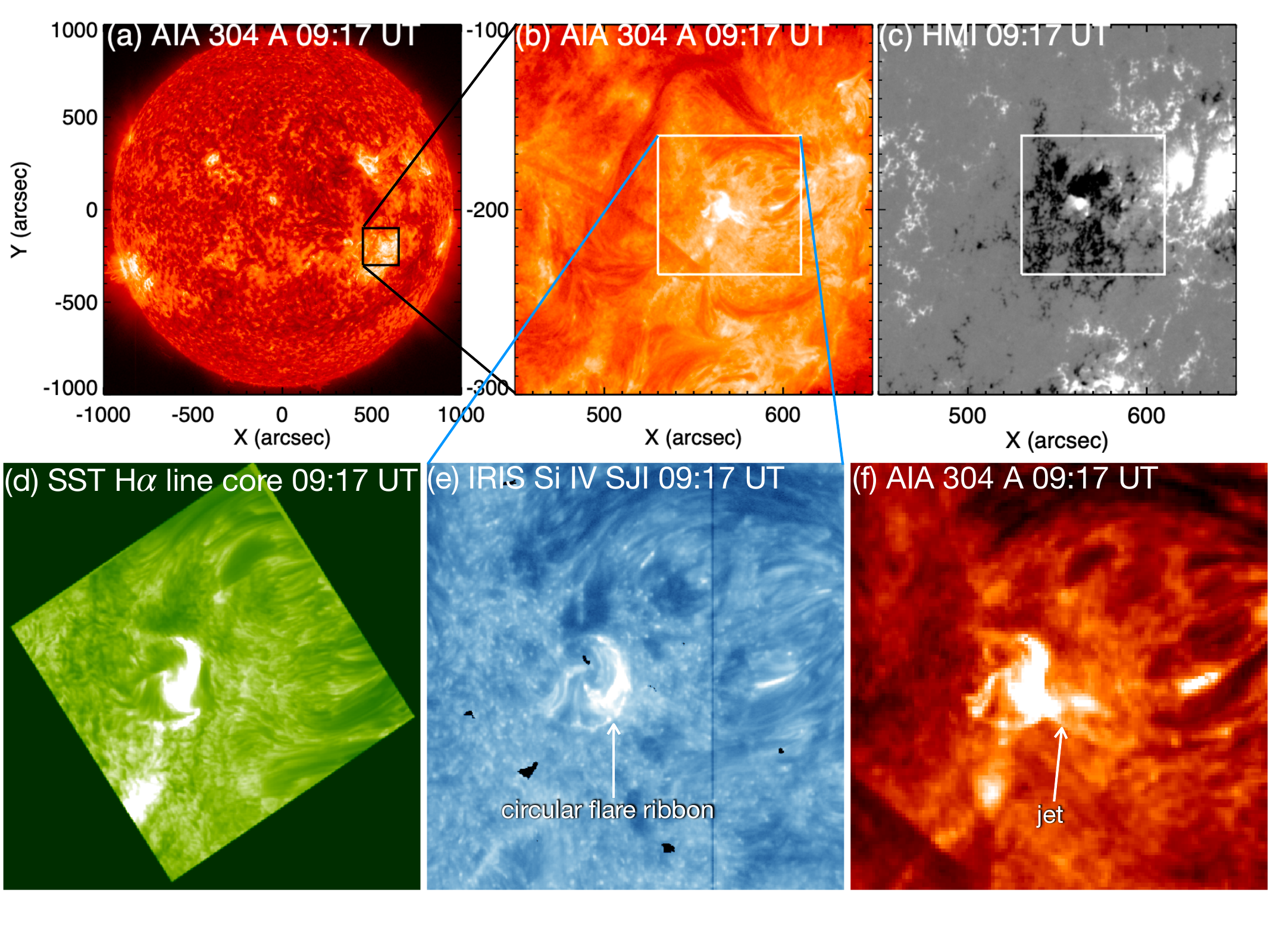}
   \caption{Multi-instrument observation of a swirled anemone jet on June 11, 2014 using SDO/AIA, SDO/HMI, SST, and IRIS. Panel (a) shows the full disk image in AIA 304~\AA, where the AR NOAA 12080 is bounded with a black square. Panel b and c are the zoomed-in views on the AR in AIA 304~\AA\ and line of sight magnetic field. The white rectangular box shows the FOV presented in the bottom row as H$\alpha$ in panel (d), Si IV in panel (e), and AIA 304 \AA.}
              \label{intro_obs}%
    \end{figure*}

\begin{figure*}[t!]
   \centering
   \includegraphics[width=\textwidth]{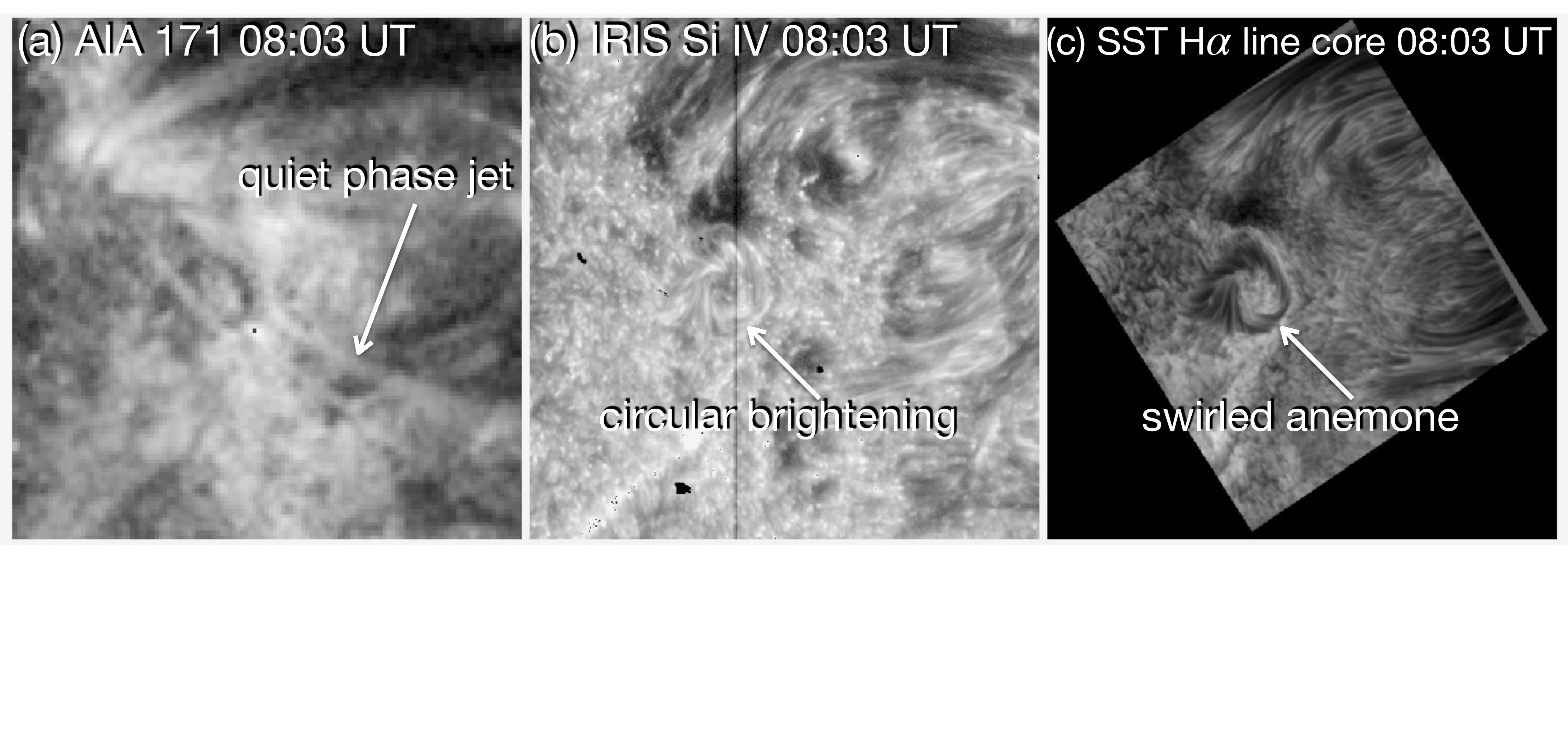}
    \vspace{-2.1cm}
   \caption{Quiet phase observations: Quiet or energy storage phase of the jet is shown in AIA 171 \AA\, IRIS Si IV SJI, and in SST H$\alpha$ observations. In hot AIA channel (panel a), a quiet phase jet spine is present at 08:03 UT (one hour before the main jet event). The circular brightening is evident in the transition region temperature observed with IRIS (panel b) and the dark absorbed structure of the swirled anemone is observed in the chromospheric H$\alpha$ observations (panel c). The vertical dark line is the IRIS SJI is the position of the slit. The FOV is presented as a white rectangular box in Fig.\ref{intro_obs} (b-c). This figure is associated with an animation, showing the evolution of the region from 08:03 UT to 09:28 UT (\url{https://www.dropbox.com/scl/fi/c1r7si5lgm4cz039u8zwv/Fig2_widejet.mp4?rlkey=5f4zahb3ybbhecwwotn6plncx&st=mscug0mf&dl=0}).}
              \label{obs1}%
    \end{figure*}
\section{Observations of the jet event}
\label{observations}
The jet was observed during a coordinated observing campaign between the Swedish 1-m Solar Telescope \citep[SST,][]{Scharmer2003} and the Interface Region Imaging
Spectrograph \citep[IRIS,][]{Pontieu2014} in active region AR12080 on June 11, 2014.  
We further use observations from the Solar Dynamics Observatory \citep[SDO,][]{Pesnell2012}: ultraviolet (UV) and extreme ultraviolet (EUV) images from the Atmosphere Imaging Assembly \citep[AIA,][]{Lemen2012}, and photospheric magnetic field maps from the Helioseismic Magnetic Imager \citep[HMI,][]{Schou2012}.
The active region was situated in the South-West part of the solar disk centered at x = 709$\arcsec$ and y = $-203\arcsec$. We focus on an area with parasitic positive magnetic polarity surrounded with negative polarity, which is the region of interest with a circular flare ribbon and jet formation  (see Fig.~\ref{intro_obs}).
The active region at the western side of the fulldisk image of the Sun in panel (a) is zoomed in panel (b) in EUV wavelength 304 \AA. 
The distribution of the magnetic field at the active region's location is depicted in panel (c) during the jet's occurrence. The lower row offers a detailed exploration of the jet region, presenting observations at various wavelengths obtained from different instruments. Details regarding the datasets utilized and the observational event are provided in the following subsections. The process of jet formation and ejection is divided into three stages: the quiet phase, the impulsive phase, and the recovery phase.
These phases are described in detail in Sects.~\ref{obs:quiet}, \ref{obs:impulsive}, and \ref{obs:recovery}, respectively. 

\subsection{Instruments}
We use the chromospheric H$\alpha$ observations from the CRisp Imaging SpectroPolarimeter \citep[CRISP;][]{Scharmer2008} instrument at the SST.
The H$\alpha$ line was sampled at 15 spectral line positions at a temporal cadence of 11.4~s and with a pixel size of 0\farcs058. 
The CRISP data was processed with the Multi-Object Multi-Frame Blind Deconvolution \citep[MOMFBD;][]{Noort2005} image restoration technique. 
In addition to the chromospheric observations from CRISP, we include \ion{Si}{iv} transition region observations from the 1400\AA\ slit-jaw (SJI) channel from IRIS. The SJI 1400 \ion{Si}{iv} images were recorded at an average temporal cadence of 17~s and a pixel size of 0\farcs166. 
The aligned SST and IRIS data were earlier analysed by
\citet{2015ApJ...809L..30C}  
and \citet{2016ApJ...817..124S},  
and are publicly available \citep{2020A&A...641A.146R}. 
The SST data were processed following an early version of the CRISPRED 
\citep{2015A&A...573A..40D} 
reduction pipeline which includes Multi-Object Multi-Frame Blind Deconvolution \citep[MOMFBD,][]{Noort2005} 
image restoration. 
For more details on the SST data processing and SST to IRIS alignment, we refer to \citet{2020A&A...641A.146R}. 
These chromospheric and transition region observations were further combined with the AIA 304\AA\ and 171\AA\ channels that provide details of the hotter components of the jet eruption. 
We use one HMI map to show the magnetic field configuration at the jet region, where a small positive polarity patch is embedded inside an extensive area with  negative magnetic polarity.
The jet region is also scanned through the IRIS slit, however, we are not using the spectral information, as this is out of scope of this paper.

\begin{figure*}
   \centering
   \includegraphics[width=0.9\textwidth]{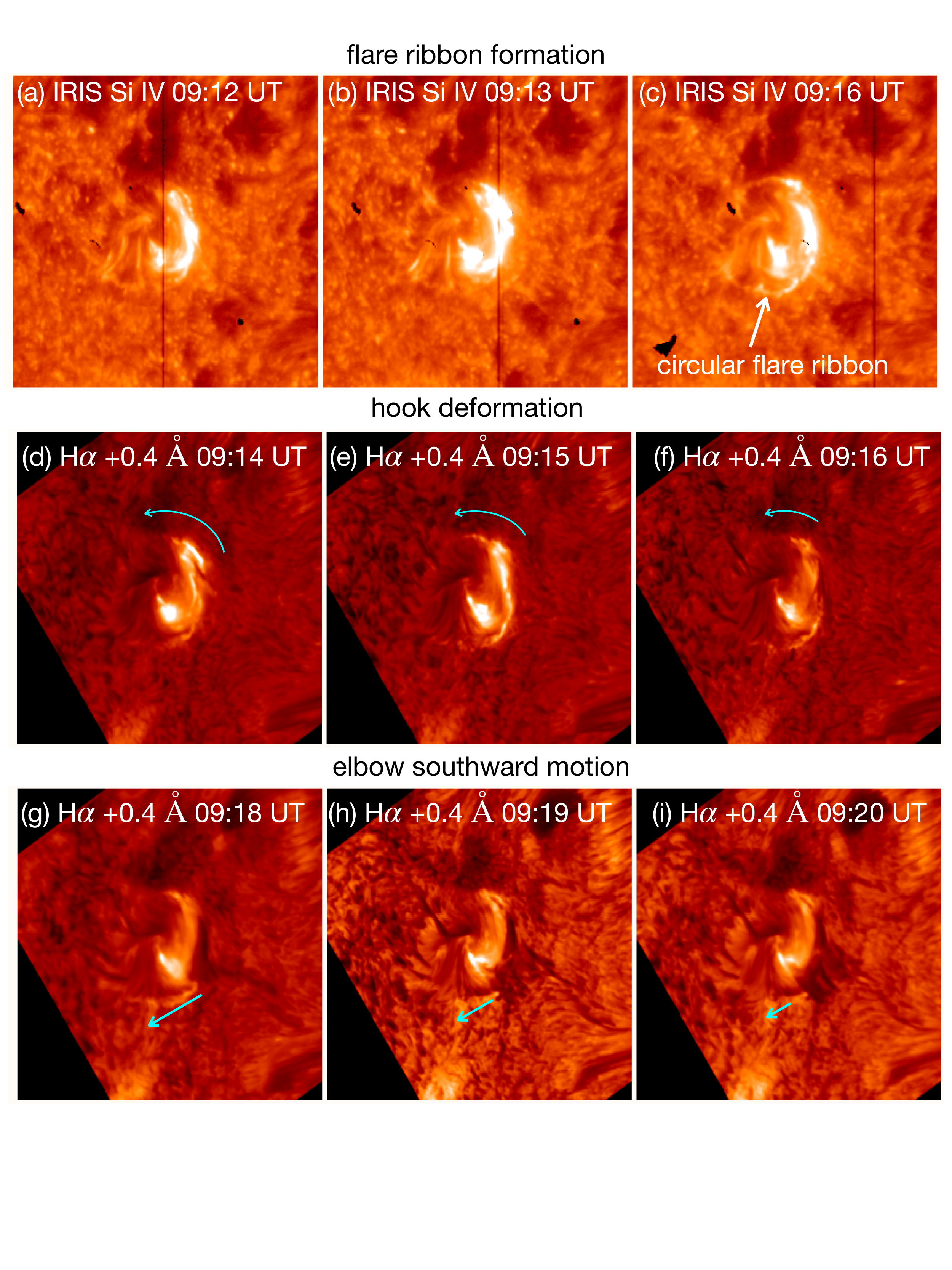}
   \vspace{-2.2cm}
   \caption{Impulsive phase observation: This energy release phase is shown in three stages. Top row: the circular flare ribbon formation in IRIS Si IV observations. Middle row: deformation of the hook brightening in H$\alpha$ red wing shown with curved cyan arrows. Bottom row: southward motion of the dark elbow-shaped structure in H$\alpha$ red wing depicted with the shortening of the arrow length.}
              \label{obs2}%
    \end{figure*}

\begin{figure*}
   \centering
   \includegraphics[width=0.9\textwidth]{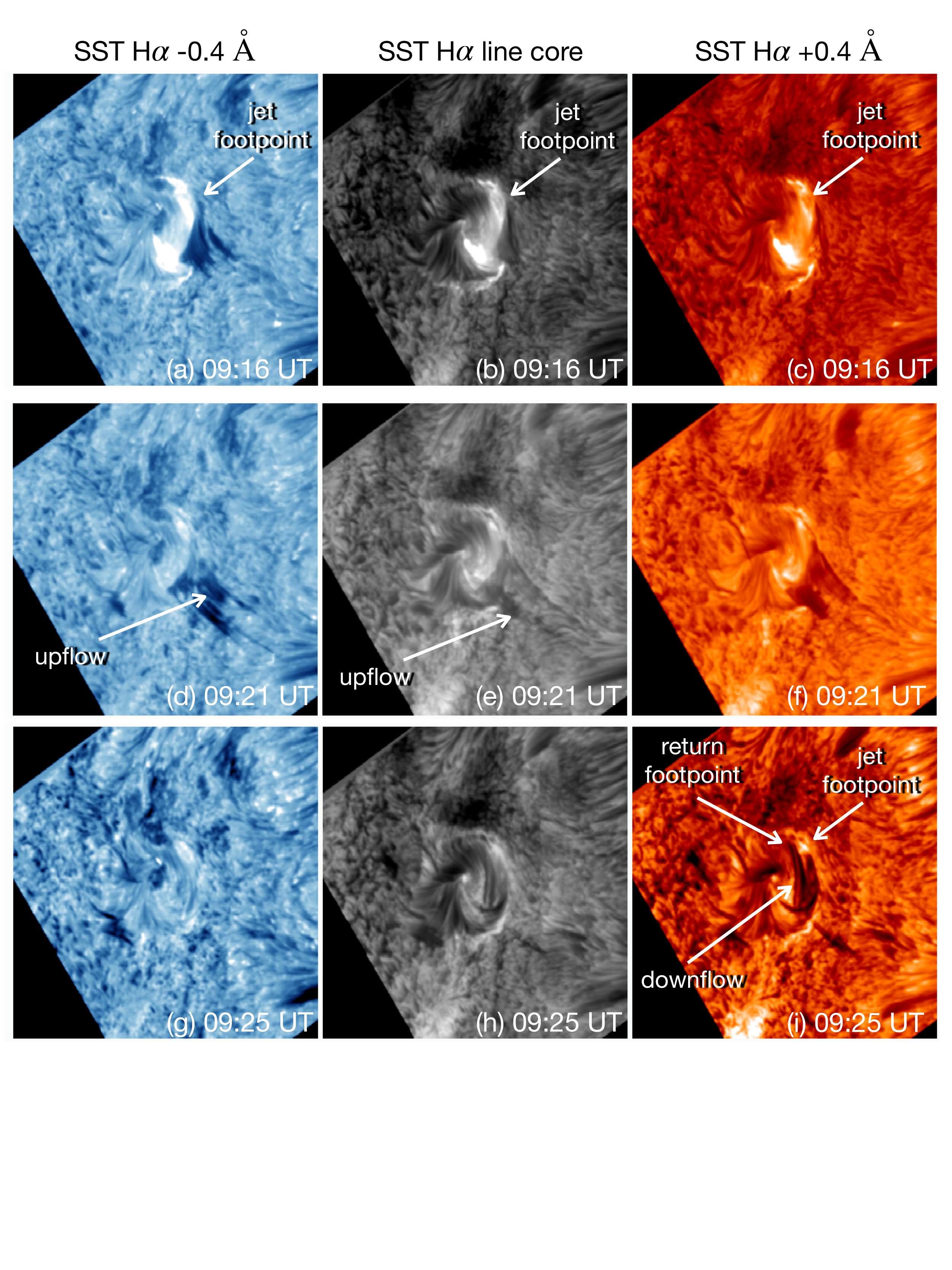}
    \vspace{-3.0cm}
   \caption{Recovery phase observation: The last phase of the jet eruption, where the up- and  downflow of the jet plasma is shown in SST H$\alpha$ blue wing (first column), line core 6563 \AA\ (middle column), and red wing (last column) observations. The initial jet footpoint is shown in first row, from where the jet starts to erupt as a wide upflow shown in panel (d-e). Some part of the jet material fell back at a different location: ``return footpoint'', shown in panel(i). }
              \label{obs3}%
    \end{figure*}

\subsection{Quiet phase} 
\label{obs:quiet}
This AR was a target for coordinated observation with SST and IRIS on June 11, 2014 starting from 07:36 UT, so we examined it for more than an hour before the jet eruption along with the circular solar flare started $\sim$09:06 UT.
At 08:03 UT, we observed a very clear swirled anemone structure in SST H$\alpha$ line core observations, and a circular brightening at the same location in transition region temperature in IRIS \ion{Si}{iv} SJIs. Examining the hot temperature EUV channel, interestingly, we observed a straight spine in AIA 171 \AA\ originating from the swirled anemone base. These observational features are presented in Fig. \ref{obs1} at 08:03 UT observed with SDO/AIA, IRIS, and SST instruments. As this straight and narrow spine in hot EUV channel is observed one hour before the main jet activity, we named it as the quiet phase jet. 
In the hot EUV channels this quite jet 
appears at 07:40 UT disappears soon after ten minutes it disappears and reappears at 08:02 UT.
So this long loop present in hot EUV channels probably corresponds to some pre-eruptive quasi-steady reconnection ongoing at that location. Its labelling as a jet may be subject to caution, although its relative intermittency may be regarded as being reminiscent of previously-reported hot narrow jets \citep[e.g. some of those seen in][]{Cirtain2007}.

\subsection{Impulsive phase}
\label{obs:impulsive}
The swirled anemone structure observed in the quiet phase started to develop a solar flare $\sim$09:05 UT and a clear circular flare ribbon formed at $\sim$ 09:16 UT around the anemone-shaped base. Along with the solar flare a wide solar jet was observed starting at $\sim$ 09:14 UT travelling towards the South-West. The circular flare ribbon formation is evident in the high resolution IRIS observations (Fig.\ref{obs2}a--c).  The jet plasma shows an upward flow in the beginning and followed by falling back jet material. From an analysis of the jet H$\alpha$ profiles, we infer Doppler offsets of about $-1.2$~\AA\ from the line core, indicating upward velocities around 55~km~s$^{-1}$ along the line of sight. Typically, velocities measured in cool jets (surges) are of this magnitude or even smaller, both from observations \citep[e.g.,][]{Schmieder1983, Schmieder1995, Joshi2020} and simulations \citep[e.g,][]{Moreno2013, Nobrega2016}.

H$\alpha$ observations in the red wing (6563 +0.4~\AA) are used to probe the downflowing of some of the jet material. With these high resolution observations, it has been found that a ``hook'' like structure is developed in the bright flare ribbon in the North-East of the anemone region  (see Fig. \ref{obs2}). This hook structure corresponds to a portion of the flare ribbon. In the standard flare model, the ribbons would be magnetically connected to the reconnection site, and would map the footpoint of separatrix field lines \citep[e.g.][]{Savcheva2015,Savcheva2016}.
A minute or two after, this hook-shaped structure starts to deform with time and fade away afterwards (Fig. \ref{obs2} middle row). We highlight this deformation of hook in Fig.~\ref{obs2} with curved cyan arrows (panels d--f) in H$\alpha$ red wing observations. The arrow of the curve is fixed at the North-East side of the hook shape, while the tail of the curve is at the head of the hook. With time, the shorter length of the cyan curve shows the deformation of this hook shape structure.

After the deformation of the hook structure, we observed the widening of the jet towards the South in a clockwise direction.  An elbow-shaped dark curtain of absorbed plasma material appeared in H$\alpha$ observations, with a sharp boundary in the South at $\sim$ 09:18 UT (see Fig. \ref{obs2}g). It appears to be moving towards the South in the clockwise direction from $\sim$ 09:18--09:23 UT. To reflect the southward motion of this dark plasma (curtain like structure), we put a straight arrow starting from the elbow of the curtain and heading towards South in Fig.~\ref{obs2}g-i. The arrow head is fixed in all three panels and the shortening of the cyan arrows depicts the widening of the jet towards the South with the elbow at the front.

\subsection{Recovery phase} 
\label{obs:recovery}
 Inspecting the three different spectral lines of the SST H$\alpha$ observations: blue wing (6563 $-$0.4~\AA), line core (6563~\AA), and red wing (6563 +0.4~\AA),  we observed that a part of the up-going jet material is falling back during 09:16--09:25 UT. The origin of the jet from the North of the anemone-shaped base is shown as ``jet footpoint'' and the wide jet in the blue wing is shown in Fig.~\ref{obs3}. Strong absorption in the blue and red wing of the H$\alpha$ wavelength reveals that the plasma material is being propelled upwards (blue-shifted) as well as downwards (red-shifted). It is not necessarily the same plasma material which is flowing up and down. It is interesting to notice that the jet material is falling back at an  offset location in the anticlockwise direction from the jet footpoint. From the observations, it gives an idea that the falling back jet material follows a different path than the upflow plasma material. Figure~\ref{obs3} illustrates this scenario with a time evolution of the jet flow in three different spectral lines next to each other, panel (i) presents the offset between jet footpoint and return footpoint.

\begin{figure}
   \centering
   \includegraphics[width=0.42\textwidth]{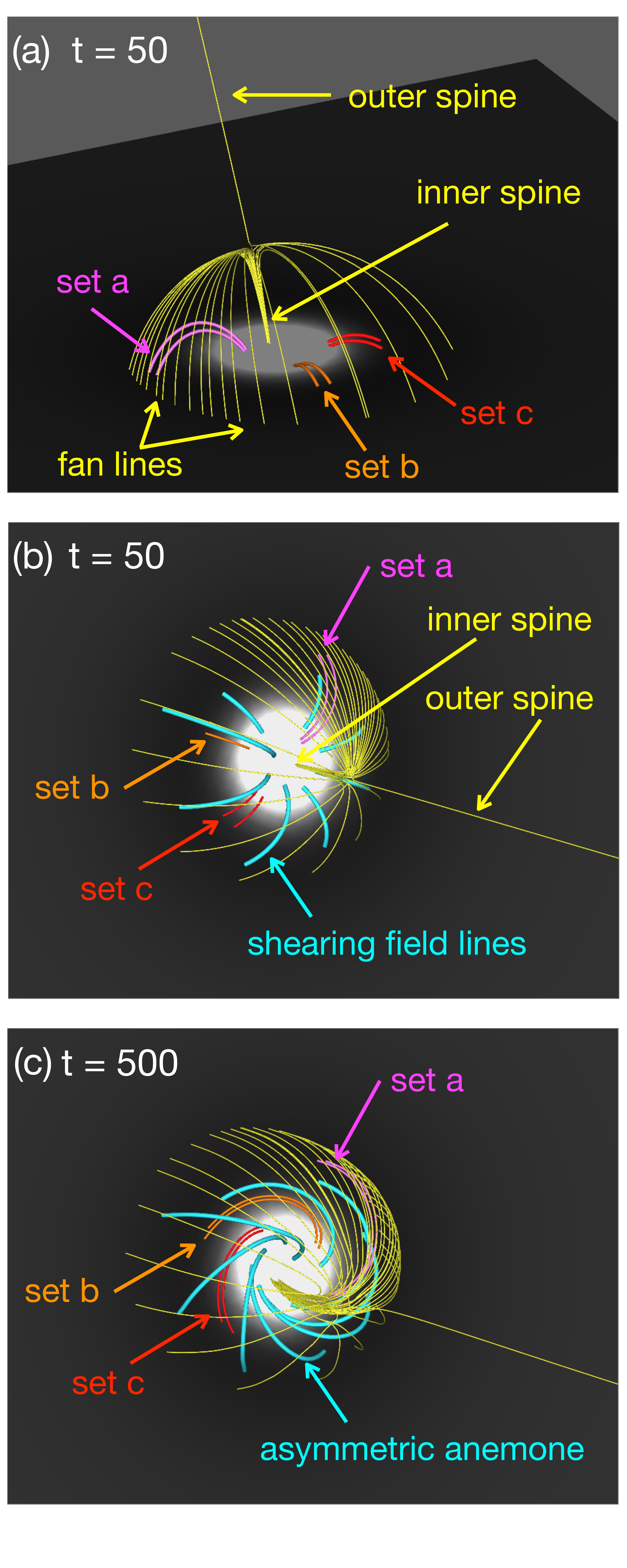}
   \vspace{-0.4cm}
   \caption{Initial configuration for the model adapted from \citet{Pariat2009} from a slanted view point (panel a), and a view point that is similar as in the observations (panel b-c). Yellow lines are grazing the separatrices, the boundary of the white circle in panel (b-c) acts as the polarity inversion line (PIL). Shearing magnetic field lines in the beginning phase (panel b) evolves as an asymmetric anemone-shape around the PIL (panel c). The different sets of field lines are explained in Sect.~\ref{sec:intro_model}.}
              \label{intro_model}%
    \end{figure}

\section{Modelling vs observed features}
\label{model}
The observed wide jet with an anemone-shaped base in the parasitic magnetic topology resembles to the generic jet model of \citet{Pariat2015} (see their Figure 1). 
In the following, we explain the main features of the original model developed by \citet{Pariat2015} and the variations we apply in this paper to that model in Sect.~\ref{sec:intro_model}, comparing the results from this numerical experiment with the observational features. 

\subsection{ARMS simulation and QSL calculation}
\label{sec:intro_model}
The model used to obtain insights on the dynamics of the observed jet is based on the numerical experiments of \citet{Pariat2015} performed with the ARMS (Adaptively Refined MHD Solver) numerical code \citep{DeVore1991}. Complete information about the ARMS code and the setup of the numerical experiment can be found in \citet[][and reference therein]{Pariat2015}. The concepts for this 3D model for solar jets were initially proposed by \citet{Pariat2009} and have been developed since then in multiple directions \citep[e.g.][]{Pariat2010,Pariat2015,Pariat2016,Dalmasse2012,Wyper2016,Wyper2018,Wyper2019,Karpen2017}. Comparisons between this model and observations have already been carried out by \citet{Patsourakos2008} in order to understand the 3D properties of coronal jets observed stereoscopically by the EUV imagers of the STEREO mission \citep{Howard2008}.

The fundamental basis of this jet model centers around the presence of a 3D null point, which partitions the coronal volume into two connectivity domains. One domain is closed, situated below the dome-like fan separatrix surface of the null point, while the other domain is open and positioned above it. In the simulations, this standard null point configuration is the result of a vertically oriented magnetic dipole, embedded slightly below the bottom boundary, which generated a photospheric magnetic field concentration, and a large scale nearly uniform and spatially slowly varying background vertical magnetic field of a direction opposite to the dipole. As can be seen in Fig.~\ref{intro_model} (a), the resulting configuration contains two distinct flux systems: a circular patch of strong closed magnetic flux surrounded by weaker open flux. Such magnetic field configuration is very commonly observed with coronal jets and coronal bright points \citep{Moreno2008,Zhang2012,Joshi2017b,Nobrega2023}.

The numerical simulation analysed in the present study is a variation the parametric simulations presented in \citet{Pariat2015}, for which the uniform background vertical field is inclined by $30^\circ$. More precisely, the magnetic field is defined by Eq. (2) of \citet{Pariat2015} with the angle $\theta$ defined by the inclination of the open field with respect to the vertical direction such as $\theta=30^\circ$ ($\theta=0$ would correspond to a vertical field). Such null point magnetic topological system first permits the very efficient storage of magnetic energy in the closed domain either thanks to shear \citep[e.g][]{Pariat2009} or the formation of a twisted flux rope \citep[e.g][]{Wyper2018}. In the present study, the injection of magnetic energy and helicity in the system follows \citet{Pariat2015} : very slow (with respect to the local Alfvén speed)  horizontal motions are applied at the line-tied photospheric boundary, only in the main central positive of the closed magnetic polarity. The system evolves quasi-steadily, with field lines being slowly sheared above the PIL (see Fig.~\ref{intro_model}, panels b and c). As the magnetic system acquires poloidal flux, the magnetic system slowly bulges and the fan dome and 3D null point rise. Similarly to \citet{Pariat2009}, the driving motions are along the isocontours of the magnetic field. However unlike with the axisymetric configuration of \citet{Pariat2009}, because of the inclination of the vertical field, the driving motions also become asymmetric. This naturally results in asymmetrically sheared field lines, as observed in Fig. \ref{intro_model}c. This is reminiscent of the observed ``swirled anemone'' feature of the present event and justify the choice of driving used here for the comparison between the observation and the numerical model (cf. also \ref{sec:SimEarlyRecon}). 

Finally, the null point configuration enables a very efficient energy release, thanks to 
magnetic reconnection \citep[e.g][]{Pariat2009,Pariat2010}, inducing the self-consistent generation of complex impulsive solar jet-like eruption, presenting both bulk flows and wave dynamics, being multi-velocity and multi-thermal, similarly to observed jets \citep{Patsourakos2008,Raouafi2016, Joshi2020FR}. The adiabatic energy equation used in the model of \citep{Pariat2015} only permits a limited understanding of the plasma emission and absorption in comparison to more sophisticated numerical experiments \citep{Fang2014,GonzalezAviles2020,ChenF2022}. Nonetheless in the low beta coronal environment, the model is fully capable of capturing the essential dynamics of magnetic field lines. Of significant importance is the implementation of an adaptive mesh refinement strategy, as detailed in the \citealp[Appendix of][]{Karpen2012}, which enhances grid resolution at the sites of reconnecting current sheets. This enhancement allows for the precise identification of diffusion regions and the reconnection site. Such an approach enables a confident comparison with observational data, as elaborated in Sect.~\ref{sec:SimEarlyRecon}

In addition, in order to analyze and visualize the magnetic topology and the evolution of the magnetic connectivity, we compute the squashing factor Q \citep{Titov2002,Titov2007}. The squashing factor is a measure of the gradients of the connectivity mapping between two planes. Numerically, the computation was done following the \citet{Aslanyan2021} implementation of the GPU-compatible QSL Squasher code \citep{Tassev2017}. More specifically, following \citet{Aslanyan2021}, we here plot a modified version of the signed log Q \citep[slog Q][]{Titov2011} initially introduced by \citet{Titov2011}, in which positive and negative values of Q are computed between different couples of reference planes. Applying the red–blue palette to the slog Q distributions, we are able to simultaneously visualize (quasi-) separatrix footprints for both open (negative/blue) and closed (positive/red) magnetic field. One shall keep in mind that since the Q values are computed between different planes for the open and closed field, the absolute Q values do not have the same significance. However since we are here only interested in the morphology of the Q distribution as well as the dynamics of the separatrix between open and closed field, the slog Q map are very instructive, in particular to understand and interpret the observed dynamics of the flare ribbons (see Sect.~\ref{sec:SimRibbon2}). 

\begin{figure*}
   \sidecaption
   \includegraphics[width=12cm]{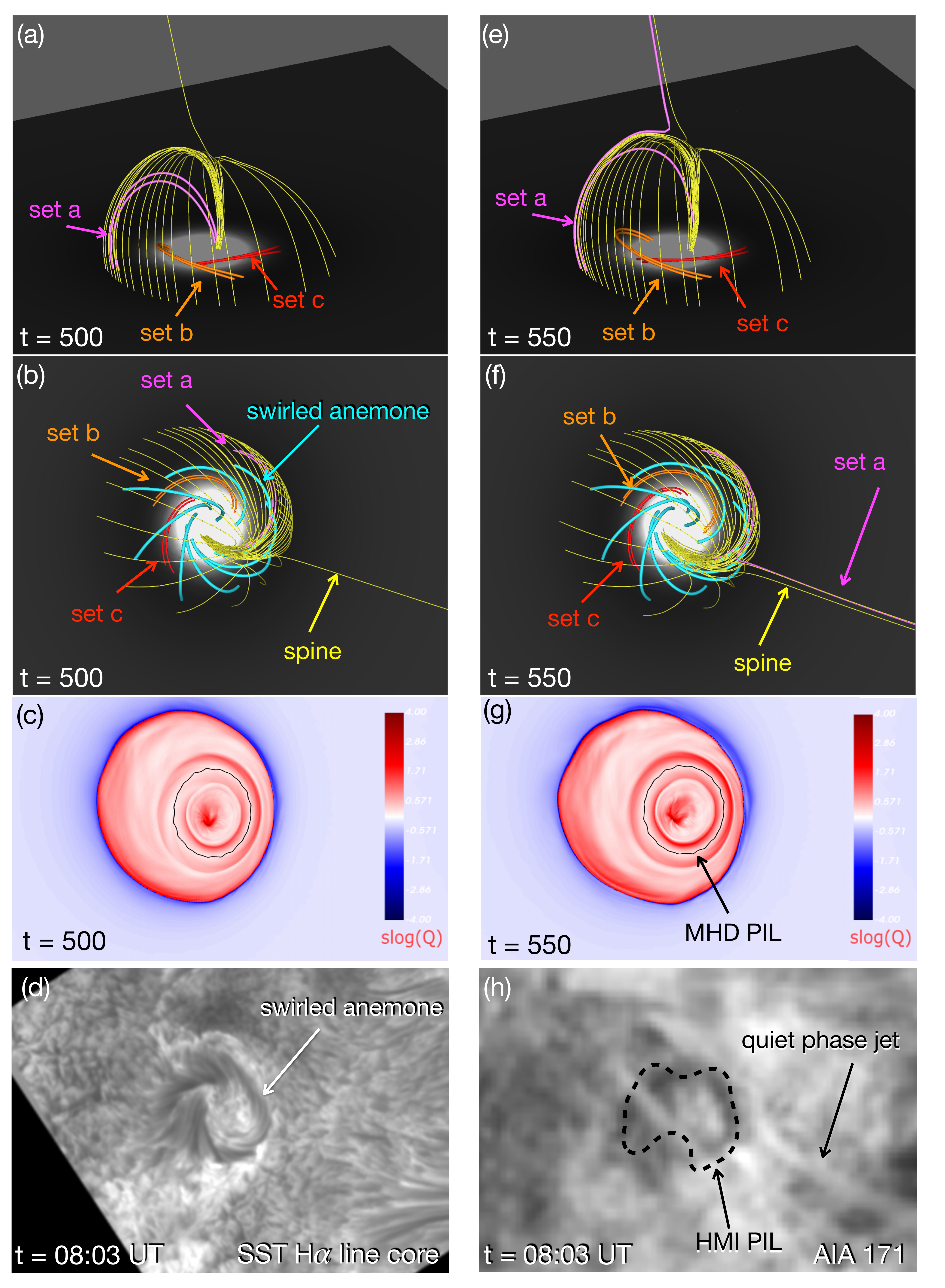}
   \caption{Modelling vs observations phase I: early reconnection phase in the ARMS simulation  (first and second row), QSL calculations (third row), and in observations (last row). First row is the field line view from side, second and third rows have the same orientation as of observations. Swirled anemone shape in the simulation is shown in panel (b) around the MHD PIL explained in the QSL map in panel (g). The similar shape in observation is shown in SST H$\alpha$ line core observations (panel d) around the HMI PIL overplotted in AIA 171 \AA\ (panel h).}
              \label{compare1}%
    \end{figure*}
\begin{figure*}
    \sidecaption
   \includegraphics[width=12cm]{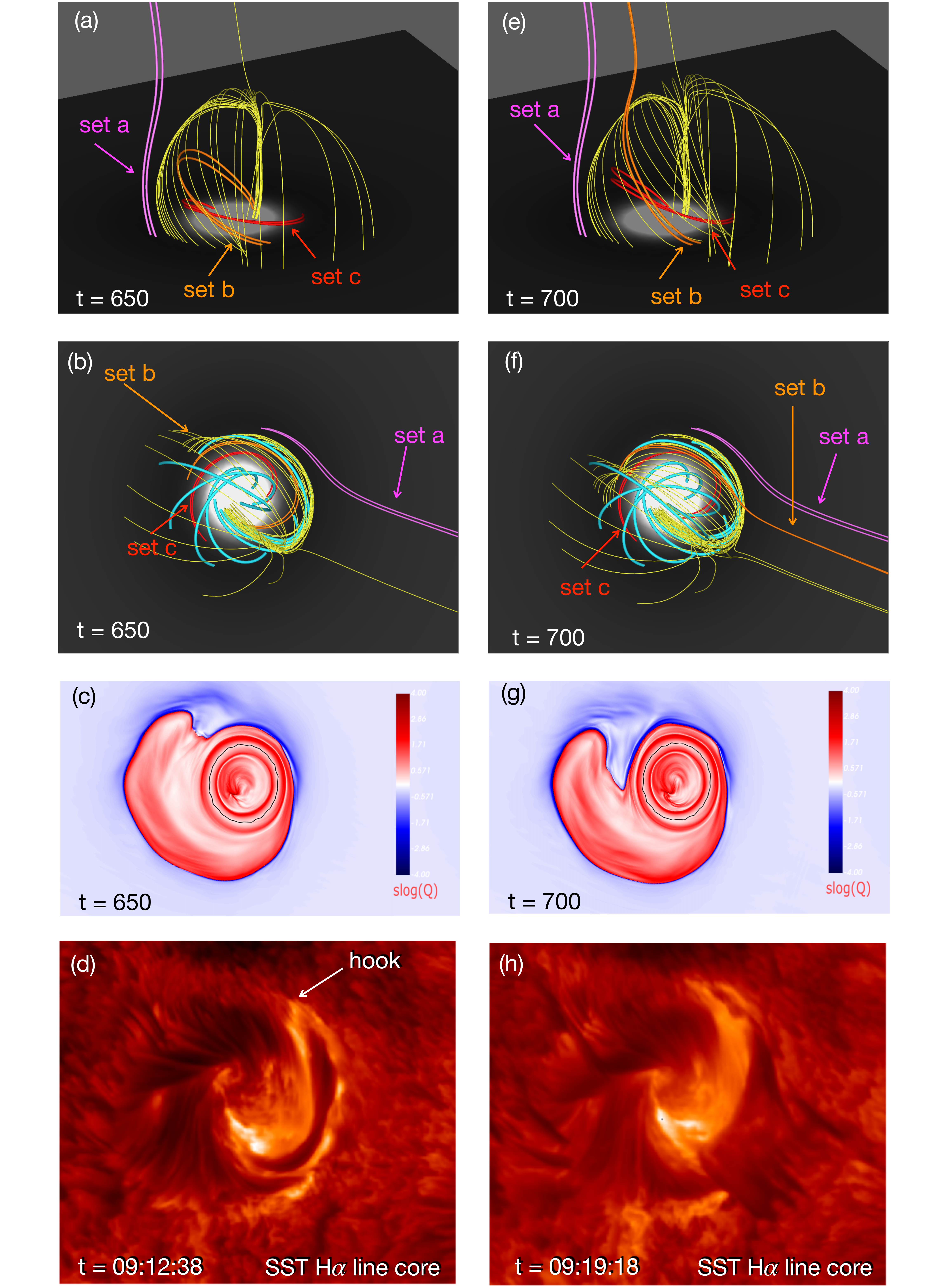}
   \caption{Modelling vs observations phase II: the impulsive phase of the jet in the field line simulation from side view (first row), observational view (second row), squashing factor maps (third row), and SST H$\alpha$ line core observations (last row).}
              \label{compare2}%
    \end{figure*}
    
\subsection{Early reconnection and straight jet}\label{sec:SimEarlyRecon}

In the relatively earlier phase of the simulation, the magnetic shear of the coronal arcades (as induced by the slow rotational boundary motion prescribed in the parasitic positive polarity) has resulted in a gradual bulging of the shearing field lines. Given the asymmetric nature of the initial magnetic field configuration, the bulging is actually not axisymmetric around in the inner spine. Indeed, simple geometrical arguments point to the fact that relatively-larger shear angles are induced in relatively-shorter field lines for the same shearing footpoint-displacements. These different shear angles generate relatively-stronger Lorentz force imbalances in relatively-shorter and more-sheared field lines, which therefore tend to expand relatively more to find their sheared equilibrium. This eventually results in one section of the fan surface (the one on the left in panel (a), being on the right in panel [b], in both Fig.~\ref{intro_model} and Fig.~\ref{compare1}) along with its corresponding ‘set a' of underlying field lines, to actually bulge more than any other part of the fan and any other set of field lines. This differential bulging is responsible for the gradual formation of a current sheet around the null point (as manifested by the acute angle made between the fan surface and the spine field line seen in Fig.~\ref{compare1}a). This gradual current-sheet formation process due to field-line bulging is actually similar to what happens in simulations of flux emergence on one side of a null point \citep[e.g.][]{Moreno2013,Nobrega2016}

In the current model, at time t = 500, the asymmetric swirled-anemone has evolved into a configuration where its sheared-most field lines (illustrated in cyan in Fig.~\ref{intro_model} and Fig.~\ref{compare1}) are now almost aligned with the polarity-inversion line, as depicted in Fig.~\ref{compare1}a-b. This occurs nearly simultaneously with the current sheet reaching the scale of the mesh in the vicinity of the coronal null-point, approximately at $t\simeq550$. This event signals the initiation of magnetic reconnection and the beginning of the jet.
Nevertheless, the resulting modeled jet is yet in a relatively quiet stage. Indeed, the reconnection and the plasma dynamics are not very impulsive. They do neither involve any large-scale restructuring of the magnetic field connectivities, nor do they lead to bulk plasma acceleration at the scale of the whole system. Instead, field lines reconnect gradually, in a quasi two dimensional way, with little total flux-transfer per unit-time from below the fan surface to above it, along the outer-spine field line. The latter can be seen qualitatively from Fig.~\ref{compare1}. Indeed, panels (e,f) show one of the pair ‘set a' field lines (drawn in pink) that has reconnected between $t=500$ and $t=550$, while all other represented field lines, including the other very close-by ‘set a' field line, have hardly evolved between these two times. Also, panels (c,g) show how the footpoints of the quasi-separatrix layer (QSL) have almost-rigidly and only-slightly shifted (toward the left) during this reconnection, sweeping a relatively small area that contains the footpoint of the single ‘set a' reconnecting field line. 

These signatures are typical of a two-dimensional-like reconnection pattern that does not involve the whole magnetic field configuration. Instead, these manifestations are, as expected, reminiscent of what was already extensively described in \citet{Pariat2015}, i.e. what they named the ‘‘straight jet'' phase, as illustrated in the left panels of their Fig.~1. Interestingly, it is also at the same time that the swirled anemone and that the straight/quiet jet are seen together in the SST and SDO observations of the event studied in this paper (see Fig.~\ref{compare1}d,h). So far, all these behaviors are typical of non-helical jets \citep[as reviewed in][]{Raouafi2016} and of some circular-ribbon and null-point related confined-flares \citep[see e.g.][]{Masson2009,Guglielmino10,Zuccarello2017a,Prasad2020,Devi2020,ChengNatCom2023}. But in the present model, they only represent the precursor of a much more dynamic phase, whose related jet was extensively studied by \citet{Pariat2015}, and which we analyze hereafter the low-altitude manifestations. 

\begin{figure*}
   \centering
   \includegraphics[width=0.9\textwidth]{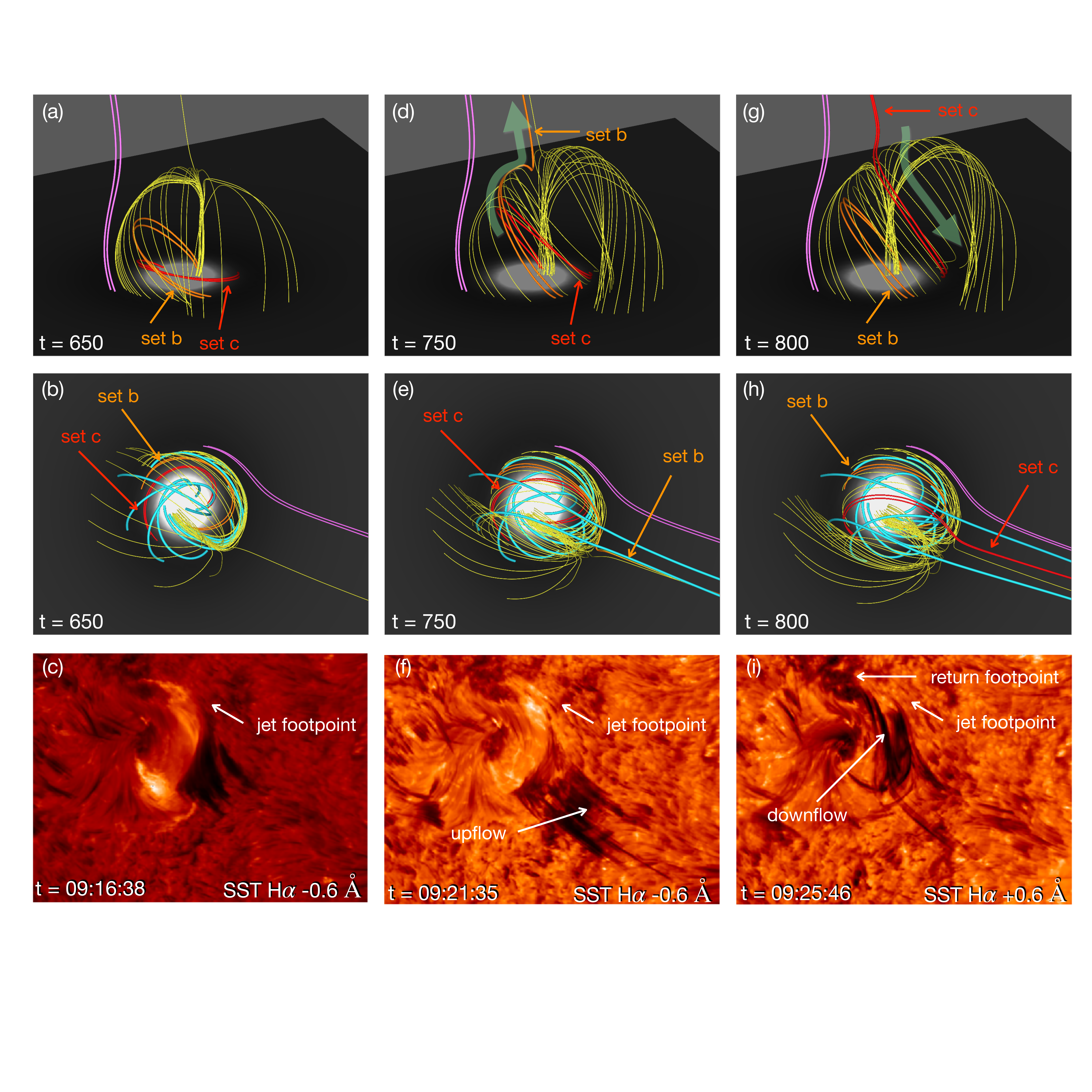}
   \vspace{-2.4cm}
   \caption{Modelling vs observations phase III: The last stage of the jet eruption, where a part of the jet material fell back on a different footpoint than the initial footpoint. First row is the side view of the simulated field lines, second row is the observational view of the simulation, and third row presents the H$\alpha$ observations in blue (panel c, f) and red (panel i) wing. Thick green arrow in panel (d) shows the direction of jet upflow along the open set b field lines, and in panel (g) shows the downflow of the jet material along the set c field lines.}
              \label{compare3}%
    \end{figure*}

\subsection{Fast widening of the jet} \label{sec:SimRibbon1}

By time $t=650-700$, the system endures a global ideal kink-instability and produces a ‘‘helical jet'', as described in \citet{Pariat2015}. Firstly, this kink instability pushes against the fan surface from below, resulting in an extension of the current sheet over a significant fraction of the fan, not only in the vicinity of the null point. Secondly, this extended current sheet provides a widespread region over which magnetic reconnection occurs, involving sheared field lines located at various places under the fan, and eventually allowing the widening of the jet over a transverse scale-length comparable to the size of the fan. This widening of the model jet is illustrated in Fig.~\ref{compare2}a--f where the the second field line of ‘set a' (in pink) has reconnected in the time interval $t=550-650$ between and both field lines of ‘set b' (in orange) are reconnecting in the time interval $t=650-700$. The distance between these field lines, as well as with the outer spine (in yellow) shows the width of the modeled jet.

Upon revisiting our SST observation, it becomes evident that the studied jet not only widened resembling a dark curtain but also extended clockwise towards the south, with its leading edge forming an elbow-like structure (see the cyan arrow in Fig.~ \ref{obs2}g). This may lead to think that the jet was extending toward the null point, being located at the top of the elbow. Such a behavior would be puzzling, since it would imply that the jet did not extend away from the null point, as post-reconnected relaxing field lines should normally behave. Instead we conjecture that the elbow of the dark curtain actually marks the position of the null point (or almost). In this line,  what SST shows would be a fast shift in position of the null toward the south during the impulsive phase of the jet. However, this shift does not happen in the MHD simulation. Oppositely, a shift in the opposite direction would even have been expected. Indeed, both the magnetic reconnection and the jet itself are removing free magnetic energy from below the fan. This energy decrease should force the magnetic configuration back towards the potential field, for which the null point is offset counterclockwise as compared to the jet state (compare Fig.~\ref{intro_model} and Fig.~\ref{compare2}). However, the counterclockwise rotation of the null point does not manifest in the simulation either. We propose that the observed stagnation of the modeled null point may be attributed to the natural counterclockwise rotation resulting from energy decrease, which must be counterbalanced by a clockwise rotation. This clockwise rotation could potentially be induced by the coupling between the kink instability and reconnection, leading to large-scale rotational motions at the helical jet's base. However, this conjecture requires further quantitative investigation in future studies.

\subsection{A hook moving along the circular ribbon} \label{sec:SimRibbon2}

Following the usual association between QSL footpoints in models and flare ribbons in observations, the simulation qualitatively retrieves the occurrence and the displacement of the observed hook like structure in the  circular flare ribbon, located at the north-west of the parasitic polarity (see Fig.~\ref{obs2} and Fig.~\ref{compare2},d,h) for the SST observation). This match is only qualitative, however. Indeed the hook in the model is rather located at north-east of the parasitic positive polarity, and even evolves toward the east (see Figure \ref{compare2},c,g). Nevertheless, it is interesting to note that this evolution of the circular ribbon is not only very different from the earlier quiet/straight-jet phase described above, but also that it was not prescribed a priori in the simulation setup. Instead, this is a behavior that spontaneously emerged from the free evolution of the system, once the photospheric driving was stopped. 

This hook-shape structure in the circular ribbon actually traces the edges of  ‘‘a channel of locally open field forms, penetrating deep into the previously closed field region as far as the polarity inversion line'', as written by \citet{Wyper2016} who obtained the very same behavior in their jet simulation, as plotted in their Figure 8. This origin and the shift in position of this finite-size inclusion of reconnected field are difficult to firmly establish. But both seem to be associated with the deformation of the flux tube that surrounds the inner spine, while it endures the ideal kink instability. This ongoing coronal deformation of the flux tube would naturally be responsible for a gradual shift in position of the bulging of the field lines below the fan, even after the photospheric drivers have been switched off, as these field lines are pushed by the kinking inner-spine. This relates in a straightforward way to to the sequential shifting of reconnection starting with the field lines of ‘set a' (Fig.~\ref{compare1}), moving to those of ‘set b' (Fig.~\ref{compare2}), and eventually involving those of ‘set c' (Fig.~\ref{compare3}).

In this context, the presence and displacement of a hook along the circular flare ribbon illustrates the gradual shift in the positions of coronal kinking, bulging, and reconnection onto the surface. These phenomena are inherent in a swirled-anemone configuration, even in its most basic, generic, initially-symmetric form in \citet{Pariat2009}. 
 Furthermore, it is important to highlight that while the moving hooked-ribbon correlates with the overarching large-scale swirling/twisting pattern of the system, it does not signify (atleast in our simulation) the existence of a drifting low-lying flux rope positioned along the PIL, as hooked ribbons typically do in the context of two-ribbon flares \citep[see][]{Aulanier2019}. Therefore, while the hook remains associated with the presence of some twist, it does not serve as a definitive signature of a flux rope.

\subsection{Downflow of the jet plasma offset from its launch site} \label{sec:SimDownflow}

The last peculiar behavior that we noticed in this jet was the falling back of some of the jet material at a ‘return footpoint' located at a different position (counterclockwise) than the ‘jet footpoint', the latter corresponding to the launch site of the material from the chromospheric layer (see Fig.~\ref{obs3}). On one hand, the simulation does not include gravity, and it uses an initially-uniform atmosphere. So it cannot be used to model, plot, and follow the lift-off followed by the fall of chromospheric/jet material. But on the other hand, one can still use the simulation to follow the spatio-temporal evolution of its field lines, and to use them to infer how some material would flow along the field, according to some gravitational pull toward the line-tied plane. This is how we proceed hereafter. 

The field line analysis is reported in Fig.~\ref{compare3}. There it can be seen that the field lines from ‘set b' (in orange) that have reconnected and opened into the jet at $t=750$ have actually reconnected a second time and have been closing down below the fan surface by $t=800$. As a consequence, any chromospheric material that could have been accelerated along these ‘set b' field lines, and having reached out into the jet at high altitude above the fan by $t=700$, have no way of falling back to their initial position along the same ‘set b' field lines for $t\le800$, because those lines are not connected to the jet anymore. Meanwhile, field lines from ‘set c' (in red) are reconnecting and opening into the jet at time $t\simeq800$. These ‘set c' field lines now provide new channels along which previously-accelerated material now has the possibility to fall back to the line-tied plane in the model, i.e. the chromosphere in the real Sun. And these new ‘set c' channels have their footpoints that are offset from those of ‘set b' field lines. When the model is oriented with angles corresponding to projection of the event as observed from Earth (see Fig.~\ref{compare3},b,e,h), one can see that this offset of the ‘set c return footpoints' is counterclockwise from the position if the ‘set b jet footpoint'.

Interestingly, similar to the observation regarding the hook, this generic and idealized model qualitatively corresponds to the observed behavior, particularly concerning the offset of the downflowing jet material. Notably, no specific configuration was imposed in the simulation to account for this behavior. It is also noteworthy that this observed offset, which may be perceived as complex, differs from the straightforward fallback of material to its point of origin, as the latter is an inherent property of a relatively simplified model.

Finally, it should be emphasized that the model allows to make a causal association between the moving hook and the offset of the downflowing material away from the jet footpoint. The fact that both are moving counterclockwise along the circular ribbon already provided a first observational hint that they were related. Thanks to the model, we can see that they are actually coupled through the magnetic reconnection in the corona: the moving hook (see Section \ref{sec:SimRibbon2}) is located on the edge of the an equally-moving open-field penetrating-region \citep[as first noticed in][]{Wyper2016}, and that is the signature at the low atmospheric level to the coronal reconnection dynamics that closes some previously-open fields (along which jet-plasma was accelerated earlier) and that opens previously-closed and distant field-lines (along which the same plasma will later fall back, remotely from its original location), leading to forming a hook and an open-field penetrating-region that shift in position.

\section{Conclusion} 
\label{results}
      
We present a high resolution observational case-study of an anemone solar jet eruption associated with a circular flare ribbon. This event was analyzed through coordinated observations obtained with SST, IRIS, and SDO, thus providing a clear picture of the low atmospheric signatures of a wide impulsive solar jet. 
We classified as three different phases of the jet  before and during the ejection: the quiet phase, the impulsive phase, and the recovery phase. The main finding of the present study is the presence of three peculiar features in the observations: a hook formation along the circular flare ribbon, a gradual widening of the jet through the motion of its kinked edge towards the magnetic reconnection site, and the falling back of some of the jet plasma at a different location than its launching site.

 The observed parasitic magnetic field topology (concentrated positive polarity surrounded by the scattered negative polarity) and H$\alpha$ observations with a wide and impulsive jet show a similarity with the model given by \citet{Pariat2015} using ARMS simulation. Therefore, we compared the above identified features with this numerical simulation. 
 The results of the model from the ARMS simulation  
was analysed in detail to get a better understanding of the physics in comparison to the observations.

 From the modelling point of view, the quiet phase jet in the observation is explained as ``early reconnection and straight jet'' and regarded as a precursor of the upcoming dynamic jet. The impulsive phase in the observations is  compared with the model with the fast widening of the jet and a hook-like structure formation inside the circular ribbon. 
The recovery phase in the observations is illustrated in comparison to the model as the opening of a different set of magnetic field lines without putting any specific changes in the simulation. This has been noticed as an offset of the downflowing jet material from its launching site in the observations. This offset downflow of the jet plasma and the hook formation in the observations as well as in the models are linked through magnetic reconnection in the corona.  The moving hook structure is the low atmospheric response to the coronal reconnection, which is evident as the reconnection process has closed the previously opened magnetic field lines from which the jet has an upflow. A different set of magnetic field lines were opened after reconnection, from which the jet material fell back to a different position as of its launch site.  In short, this generic, and non-specific simulation reproduces the very distinct behaviours that are present in the high-resolution SST and IRIS observations. Furthermore we regard it likely that a model that incorporates chromospheric conditions will also explain  the ejection of cool material, as the anemone type structure is favourable for the ejection of cool jets.

In this study, we suggest that the components of the \citet{Pariat2015} model naturally give rise to the observed features, while acknowledging the potential for alternative explanations from different numerical experiments \citep[e.g.,][]{Archontis2013, Wyper2017}.
Here, we explain the coherence between the observed characteristics and the pre-jet physical elements in the corona. We conjecture that the highly resolved observations of a swirled anemone jet presented in this example likely reveal generic features inherent in wide and impulsive jets. We propose that their absence from earlier reports may be attributed to their potential blending with other features within more complex environments. Therefore, to detect and study such features, high-resolution observations of jets in both the solar corona and lower solar atmospheric layers using advanced imaging and spectroscopic techniques are necessary, such as: the Daniel K. Inouye Solar Telescope \citep[DKIST;][]{DKIST2020}, the Solar-C~\citep[EUVST;][]{EUVST2020}, the Multi-slit Solar Explorer \citep[MUSE;][]{MUSE2022} and the European Solar Telescope \citep[EST;][]{EST2022}.

\begin{acknowledgements}
This research has been supported by the Research Council of Norway, project number 325491, 
and through its Centres of Excellence scheme, project
number 262622, 
by the European Research Council through the Synergy Grant number 810218 (``The Whole Sun'', ERC-2018-SyG), 
by the APR program of the French national space agency (CNES), 
and by the Programme National Soleil Terre (PNST) of the CNRS/INSU also co-funded by CNES and CEA. 
This work benefited from discussions at the International Space Science Institute (ISSI) in Bern, through project \#535 ``Unraveling surges: a joint perspective from numerical models, observations, and machine learning''.
The Swedish 1-m Solar Telescope is operated on the island of La Palma by the Institute for Solar Physics of Stockholm University in the Spanish Observatorio del Roque de Los Muchachos of the Instituto de Astrof\'isica de Canarias. The Institute for Solar Physics is supported by a grant for research infrastructures of national importance from the Swedish Research Council (registration number 2017-00625). 
SDO observations are courtesy of NASA/SDO and the AIA, EVE, and HMI science teams.
IRIS is a NASA small explorer mission developed and operated by LMSAL with mission operations executed at NASA Ames Research center and major contributions to downlink communications funded by ESA and the Norwegian Space Agency.
The numerical simulations were carried on the HPC resources of the National Computer Center for Higher Education (CINES), through time allocation A0010406331 granted by Grand \'Equipement National de Calcul Intensif (GENCI). This project was also provided with computer and storage resources by GENCI at IDRIS thanks to the grant A0130406331 on the supercomputer Jean Zay's the CSL partition.
We made use of NASA’s Astrophysics Data System Bibliographic Service.
We thank to Dr. Valentin Aslanyan for making available his QSL computation routines.
\end{acknowledgements}

\bibliography{references}
\bibliographystyle{aa}
\end{document}